\documentclass[twocolumn,showpacs,showkeys,preprintnumbers,amsmath,amssymb]{revtex4}

\usepackage{hyperref}
\usepackage{revsymb}
\usepackage{graphicx}
\usepackage{dcolumn}
\usepackage{bm}

\begin{document}

\title{Controllable operation for distant qubits in a two-dimensional quantum network}
\author{Zhi-Rong Zhong$^1$}
\email{zhirz@fzu.edu.cn}
\author{Xiu Lin$^{1,2}$}
\author{Bin Zhang$^1$}
\author{Wan-Jun Su$^1$}
\author{Zhen-Biao Yang$^{1,3}$}
\email{zbyangmk@gmail.com}
\address{1.Department of Physics, Fuzhou University, Fuzhou 350002, P. R. China\\
2.School of Physics and Optoelectronics Technology, Fujian Normal \\
University, Fuzhou 350007, P. R. China\\
3.Key Laboratory of Quantum Information,University of Science and
Technology of China, CAS, Hefei 230026, P. R.China}

\date{\today }

\date{\today}

\begin{abstract}

We propose a theoretical scheme to realize the coherent coupling of
multiple atoms in a quantum network which is composed of a
two-dimensional (2D) array of coupled cavities. In the scheme, the
pairing off-resonant Raman transitions of different atoms, induced
by the cavity modes and external fields, can lead to selective
coupling between arbitrary atoms trapped in separated cavities.
Based on this physical mechanism, quantum gates between any pair of
qubits and parallel two-qubit operations can be performed in the 2D
system. The scheme provides a new perspective for coherent manipulation of
quantum systems in 2D quantum networks.

\end{abstract}

\maketitle

Coherent manipulation of quantum systems at a distance is
one of the crucial ingredients in the upcoming area of quantum technologies.
The realization of quantum networks composed of many nodes and channels provide
opportunities for such purpose, and thus is of great consequence to a series of
frontiers, such as quantum computation, communication and metrology \cite{Kimble}.
Fundamental to quantum networks are quantum interconnects, which achieve
reversible quantum state transformation among separate systems across network nodes.
Such quantum connectivity in quantum networks can be realized by matter-light interaction in cavity quantum electrodynamics setups
at nodes \cite{Haroche}, combined with channels of light tunneling across the nodes \cite{DiVincenzo}. As kinds of matter particles
(say, atoms, quantum dots, nitrogen-vacancy (NV) centres in diamond, superconducting qubits in superconducting cavity,  etc), distributed in network nodes, can act as stationary
qubits for information storage; while single photons are suitable to act as flying qubits, that are convenient for conveying
from and to the nodes the information exchanged through the matter-light interaction.

In recent years, much attention has been paid to using the atom-light interaction in coupled cavity arrays for investigating
novel physical phenomena and its possible applications \cite{Hartmann}. Many theoretical studies concerning the use of coupled cavity arrays have been done, as to realize controllable operation for
quantum simulation of many-body phenomena \cite{MJHartmann,Greentree,Angelakis,Bose,Irish,Zhou,Chen, 012307} and
for distributed quantum information processing \cite{Cirac,Pellizzari,Serafini,Yin,Songjie,Ye,XYL,Zheng,SBZheng,ZBYang,Song,JCho,LanZhou,Li,Gia,Ke,Songxiahe,ZhengNori}. All such researches focus on the cases of either two-site \cite{Wumingliu,Irish,Pellizzari,Serafini,Yin,Songjie,Ye,XYL,Zheng,SBZheng,ZBYang,Song,JCho,Gia,Ke} or one-dimensional (1D) \cite{MJHartmann,Greentree,Angelakis,Bose,Zhou,Chen,LanZhou,Li,Songxiahe,ZhengNori} coupled cavity arrays. Extending such studies to two- (2D) or three-dimensional (3D) coupled cavity arrays is in some sense of more significance, as it is shown that some kinds of 2D and 3D quantum states (say, cluster states) are universal resources for quantum computation \cite{Raussendorf}. There have been studies considering the 2D coupled cavity arrays, the correlative examples can be found in Ref. \cite{Cho} and \cite{Lin}, which respectively consider the realization of the fractional quantum Hall system and 2D one-way quantum computation. In these schemes \cite{Cho,Lin}, the nearest-neighbor coupling is essential for the nonlocal quantum coherence across the 2D network nodes.

In this paper, we first propose a scheme to control coherently the coupling
between two arbitrary atomic qubits at distant (not necessarily the nearest-neighbor)
nodes in a 2D array of coupled cavities. In the scheme, the off-resonant Raman transitions between
two ground states of the atoms, induced by the cavity modes
and the external fields, can lead to selective coupling between
arbitrary two distant atoms across the 2D network nodes, while with all the atoms as well as with all the cavity modes only virtually excited.
Quantum logic operations between any pair of distant atomic qubits and parallel two-qubit operations on selective qubit pairs can be implemented by appropriately selecting the parameters of the external fields.

We consider a coupled $N\times N$ cavity array, as shown in FIG. 1. In each site (denoted here by $jk$) there is a cavity mode,
which respectively couple to their neighboring ones through the $x$ and $y$ directions with intercavity photon hopping. Each site contains a $\Lambda $-type atom, with two ground states $\left| g\right\rangle _{jk}$
and $\left| f\right\rangle _{jk}$, and one excited state $\left| e\right\rangle _{jk}$, as shown in FIG. 2. The atom interacts with the cavity mode through the transition $\left| g\right\rangle _{jk}\leftrightarrow \left| e\right\rangle _{jk}$, with coupling rate $g_{jk}$ and detunings $\Delta_{1jk}$. A classical field with Rabi frequencies $\Omega_{jk}$ is  applied to drive the atomic transition $\left| f\right\rangle _{jk}\leftrightarrow \left| e\right\rangle _{jk}$, with detunings $\Delta_{2jk}$. In the interaction picture, the Hamiltonian for the system is

\begin{eqnarray}
H=H_1+H_2,
\end{eqnarray}
where $H_1$ and $H_2$ denote the coupling between the atoms and the field as well as between the cavities, respectively.
\begin{eqnarray}
H_1=\sum_{j,k=1}^N\ [g_{jk}a_{jk}|e\rangle _{jk}\langle g|e^{i\triangle _{1jk}t}+\Omega _{jk}|e\rangle _{jk}\langle f|e^{i\triangle _{2jk}t}\nonumber \\+h.c],
\end{eqnarray}
\begin{eqnarray}
H_2=\sum_{j,k=1}^N[\ va_{jk}a_{j+1,k}^{+}+va_{jk}a_{j,k+1}^{+}+h.c],
\end{eqnarray}
 where $a_{jk}$ denotes the annihilation operator for the mode of the $jk$th cavity, $v$ is the hopping rates of photons between neighboring
cavities.

\begin{figure}
\centering
\includegraphics[width=2.6in]{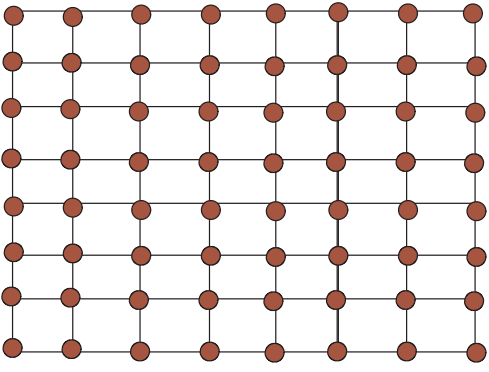}
 \caption{(Color online) Schematic diagram of a two-dimensional (2D) array of coupled cavities. Each atom is confined in the center of cavity.}
\end{figure}
\begin{figure}
\centering
\includegraphics[width=2.6in]{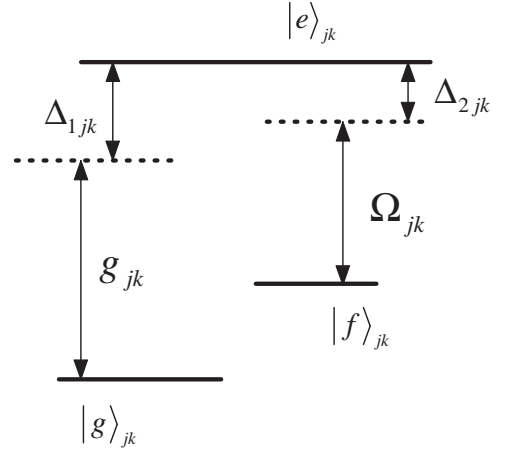}
 \caption{The atom level scheme. The transition of the $jk$th atom $|g\rangle _{jk}$ $\leftrightarrow $ $|e\rangle _{jk}$ is coupled to
the cavity mode with detuning $\Delta_{1jk}$, the corresponding coupling rate is $%
g_{jk}$. The transition $|f\rangle _{jk}$ $%
\leftrightarrow $ $|e\rangle _{jk}$ of the $jk$th atom is driven by a classical laser field with the detuning $\Delta _{2jk}$, and the corresponding the Rabi frequencies is $%
\Omega _{jk}$.}
\end{figure}

We adopt periodic boundary conditions $a_{j1}=a_{jN}$ and $a_{1k}=a_{Nk}$,
by introducing the nonlocal bosonic modes $c_{mn}$, and making the transformation
$a_{jk}=\frac 1N\sum_{m,n}^N\exp [-i (\frac{2\pi jm}N+\frac{2\pi kn}N)]
c_{mn}$. Thus we can rewrite the Hamiltonian $H_1$ and $H_2$
as
\begin{eqnarray}
H_1 &=&\sum_{j,k}\ [\Omega _{jk}e^{i\triangle _{2jk}t}|e\rangle
_{jk}\langle f|+\sum_{m,n}\frac{g_{jk}}N \nonumber\\&&\times e^{-i(\frac{2\pi
jm}N+\frac{2\pi kn} N)} e^{i\triangle
_{1jk}t}|e\rangle _{jk}\langle g|c_{mn}+h.c.],
\end{eqnarray}
and
\begin{equation}
H_2=\ \sum_{m,n}^N\omega_{mn} c_{mn}^{+}c_{mn}+h.c.,
\end{equation}
where $\omega _{mn} =2v\cos (\frac{2\pi n}N+\frac{2\pi m}N)$.
We now go into a new frame by defining $H_1$ as a free Hamiltonian, and
obtain the interaction Hamiltonian for the whole system as
\begin{eqnarray}
H_1^{^{\prime }} &=&\sum_{j,k}^N\ \{\Omega _{jk}e^{i\triangle
_{jk,2}t}|e\rangle _{jk}\langle f|\nonumber \\&&
+\sum_{m,n}[\frac{g_{jk}}Ne^{-i(\frac{%
2\pi jm}N+\frac{2\pi kn}N)}\ e^{i(\triangle _{1jk}-\omega _{mn})t}
\nonumber
\\ && \times c_{m,n}|e\rangle _{jk}\langle g|+h.c.]\}.\
\end{eqnarray}

Considering the large detuning case with $\left| \triangle _{2jk}\right|
\gg $ $\left| \Omega _{jk}\right|$, and $ \left| \triangle _{1jk}-\omega
_{mn}\right| \gg \left| \frac{g_{jk}}N\right| $, we can adiabatically eliminate the atomic excited state $%
|e\rangle_{jk} $ and turn $H_2^{^{\prime }}$ to
\begin{eqnarray}
H_1^{^{^{\prime \prime }}} &=&-\sum_{j,k}\ \ \{\ \varepsilon_{jk}
|f\rangle _{jk}\langle f|+\sum_{m,n}[\zeta _{jk}|g\rangle _{jk}\langle
g|c_{mn}^{+}c_{mn} \nonumber \\
&&+\lambda _{jk}e^{-i(\frac{2\pi jm}N+\frac{2\pi kn}N%
)}e^{i(\triangle _{1jk}-\omega _{mn}-\triangle _{2jk})t}c_{mn}S_{jk}^{+} \nonumber
\\&& + h.c.]\},
\end{eqnarray}
where $\varepsilon _{jk}=\frac{(\Omega _{jk})^2}{\Delta _{2jk}}$, $\zeta_{jk}=\frac{(g_{jk})^2}{N^2(\Delta _{1jk}-\omega _{mn})}$, $\lambda _{jk}=\frac{g_{jk}\Omega _{jk}}{2N}(\frac 1{\Delta _{1jk}-\omega _{mn}}+\frac 1{\Delta _{2jk}})$, and $S_{jk}^{+}=\left| f\right\rangle _{jk}\left\langle g\right| $.
The first and second terms of Eq.(7) describe Stark shifts respectively induced by the classical fields and bosonic modes,
while the last two terms describe the multiple off-resonant Raman transitions for each atom induced by the classical fields and the bosonic modes.  Under the condition $\left| \triangle _{1jk}-\omega _{mn}-\triangle _{2jk}\right| \gg \lambda _{jk}e^{-i(\frac{2\pi jm}N+\frac{2\pi kn}N)}$, the bosonic modes are only virtually excited. This thus leads to
the quanta-dependent Stark shifts and effective coupling between the atoms, and gives the effective Hamiltonian \cite{Zhengguo,Imamoglu}
\begin{eqnarray}
H_{e} &=&\sum_{j,k}\{\ -\varepsilon _{jk}|f\rangle _{jk}\langle f|
+\sum_{m,n}[-\zeta_{jk}|g\rangle _{jk}\langle g|c_{mn}^{+}c_{mn} \nonumber \\
&& + \xi _{jkmn}(c_{mn}c_{mn}^{+}|f\rangle _{_{jk}}\langle
f|-c_{mn}^{+}c_{mn}|g\rangle _{jk}\langle g|)] \nonumber\\
&&+\ \sum_{p,q(pq\neq jk)}\chi _{jkpq}S_{pq}^{+}S_{jk}^{-} \nonumber \\
&&\times e^{-i[(\triangle _{1pq}-\triangle _{1jk})-(\triangle _{2pq}-\triangle _{2jk})]t}+h.c\},
\end{eqnarray}
with
\begin{equation}
\xi _{jkmn}=\frac{(\lambda
_{jk})^2}{\triangle _{1jk}-\omega _{mn}-\triangle
_{2jk}},
\end{equation}
and
\begin{eqnarray}
\chi _{jkpq} &=& \sum_{m,n}\frac 12\lambda _{jk}\lambda _{pq}(\frac
1{\triangle _{1jk}-\omega _{mn}-\triangle _{2jk}}\nonumber\\&&+\frac
1{\triangle _{1pq}-\omega _{mn}-\triangle
_{2pq}})\times \nonumber\\&&e^{-i[2(j-p)\frac{m\pi }N+2(k-q)\frac{n\pi }N]}.
\end{eqnarray}
As the quantum number of the bosonic modes is conserved during the interaction, they will remain
in the vacuum state if they are initially in the vacuum state. Then $H_{e}$ reduces to
\begin{eqnarray}
H_{e}&=&\sum_{j,k}\ \{\ \varsigma_1 |f\rangle
_{jk}\langle f|+\ \sum_{p,q(pq\neq jk)}[\chi_{jkpq}S_{pq}^{+}S_{jk}^{-}\nonumber\\
&&\times e^{-i[(\triangle _{1pq}-\triangle _{1jk})-(\triangle _{2pq}-\triangle _{2jk})]t}+h.c]\},
\end{eqnarray}
where $\varsigma_1=\sum_{m,n}\xi _{jkmn}-\varepsilon _{jk}$.

The Hamiltonian in Eq. (11) allows for the coherent operation of two arbitrary distant qubits
across the 2D quantum network. In order to do so, we apply classical fields to the $jk$th and $pq$th qubits, and select the frequencies for the cavity and
classical fields in such a way that the conditions $g_{jk}=g_{pq}$, $\Omega_{jk}=\Omega_{pq}$, and  $\triangle_{1pq}-\triangle_{1jk}$ = $\triangle_{2pq}-\triangle_{2jk}$ are also fulfilled.
Thus the effective Hamiltonian $H_{e}$ reduces to
\begin{equation}
H_{e,jkpq}=\ \varsigma_1 (|f\rangle _{jk}\langle f|+|f\rangle
_{pq}\langle f|)+\ \chi_1(S_{pq}^{+}S_{jk}^{-}+h.c),
\end{equation}
with $\chi_1=\chi_{jkpq}$, and $jk\neq pq$.
We assume the $jk$th and $pq$th qubits are first in state
$|\psi (0)\rangle =|f\rangle_{jk} |g\rangle_{pq} ,$ the time evolution can be expressed as
$|\psi (t)\rangle =e^{-iH_{e,jkpq}t}|\psi (0)\rangle=e^{-i\varsigma_1 t}[\cos (\chi_1 t)|f\rangle_{jk} |g\rangle_{pq}-i\sin (\chi_1 t)|g\rangle_{jk} |f\rangle_{pq}]$. After an interaction time $\frac \pi {4\chi_1},$ the two qubits evolve to a maximal
entangled state $|\psi \rangle_{en}=e^{-i\frac {\varsigma_1\pi} {4\chi_1}}(|f\rangle_{jk} |g\rangle_{pq} -i|g\rangle_{jk}|f\rangle_{pq} )/\sqrt{2}.$
Or, if the two qubits are in state $|\psi' (0)\rangle =(c_0|f\rangle_{jk}+c_1|g\rangle_{jk} )|g\rangle_{pq}$, with $\left| c_0\right| ^2+\left| c_1\right| ^2=1$. After an interaction time $t=$
$\frac \pi {\chi_1}$, the state of the $jk$th qubit is now transferred to that of the $pq$th qubit, i.e., the two qubits are finally in
$|\psi \rangle_{st}=|g\rangle_{jk} (-ie^{-i\frac {\varsigma_1\pi} {\chi}}c_0|f\rangle_{pq}+c_1|g\rangle_{pq})$.

We notice that the selective parallel two-qubit operation on different qubit pairs can also been implemented in such a 2D model.
Suppose that one wants to perform gates on qubit pairs $(jk,pq)$ and $(j^{\prime }k^{\prime },p^{\prime }q^{\prime }).$ Then we drive each of these qubits with two classical fields. The parameters of the system are suitably adjusted so that all such conditions
$g_{jk}=g_{pq}$, $\Omega_{jk}=\Omega_{pq}$, $\triangle_{1jk}=\triangle_{1pq}$, $\triangle_{2jk}=\triangle_{2pq}$,
$g_{j^{\prime }k^{\prime }}=g_{p^{\prime }q^{\prime }}$, $\Omega_{j^{\prime }k^{\prime }}=\Omega_{p^{\prime }q^{\prime }}$, $\triangle_{1j^{\prime }k^{\prime }}=\triangle_{1p^{\prime }q^{\prime }}$, $\triangle_{2j^{\prime }k^{\prime }}=\triangle_{2p^{\prime }q^{\prime }}$, and $\left| \Delta _{1\alpha \beta } -\Delta _{2\alpha \beta } -\Delta _{1\alpha ^{\prime }\beta ^{\prime }} +\Delta _{2\alpha ^{\prime }\beta ^{\prime }} \right| \gg \chi _{\alpha \beta \alpha ^{\prime }\beta ^{\prime }} (\alpha \beta =jk,pq;\alpha ^{\prime }\beta ^{\prime }=j^{\prime }k^{\prime },p^{\prime }q^{\prime })$ are also satisfied. In such a case, qubit $jk$ ($j^{\prime }k^{\prime }$) only couples to qubit $pq$ ($p^{\prime }q^{\prime }$), while it decouples to qubits $j^{\prime }k^{\prime }$ ($jk$) and $p^{\prime }q^{\prime }$ ($pq$). Therefore the effective Hamiltonian is given by
\begin{eqnarray}
H_{e,par}&=&\ \varsigma_1 (|f\rangle _{jk}\langle f|+|f\rangle
_{pq}\langle f|)+\ \chi_1(S_{pq}^{+}S_{jk}^{-}+h.c) \nonumber\\
&&+\ \varsigma_2 (|f\rangle _{j^{\prime }k^{\prime }}\langle f|+|f\rangle_{p^{\prime }q^{\prime }}\langle f|)+\ \chi_2(S_{p^{\prime }q^{\prime }}^{+}S_{j^{\prime }k^{\prime }}^{-} \nonumber\\
&&+h.c),
\end{eqnarray}
where $\varsigma_2=\sum_{m,n}\xi _{j^{\prime }k^{\prime }mn}-\varepsilon _{j^{\prime }k^{\prime }}$, $\chi_2=\chi_{j^{\prime }k^{\prime }p^{\prime }q^{\prime }}$, and $j^{\prime }k^{\prime }\neq p^{\prime }q^{\prime }$.
This thus allows us to coherently perform two-qubit operations on qubit pairs ($jk,pq$) and ($j^{\prime }k^{\prime },p^{\prime }q^{\prime }$) simultaneously.

\begin{figure}
\centering
\includegraphics[width=3.6in]{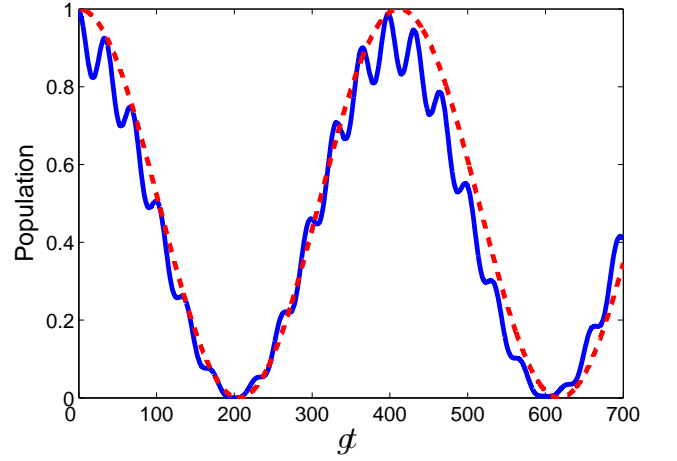}
 \caption{(Color online) Numerical illustration for the time evolution of the occupation probability $P=\left| \left\langle \psi |\psi(t) \right\rangle \right| ^2$ governed respectively by the effective (blue-solid line) and full (red-dashed line) Hamiltonian models, the atoms are initially in state $\left| f_{_{11}}g_{_{12}}\cdots g_{jk}\cdots g_{44}\right\rangle $, while all the cavity modes are in vacuum state. The relative parameters are set to be :$v=1.5g$, $\Omega _{11}=\Omega _{44}=1.0g$, $\Omega _{ij}=0 (i,j=1,2,3,4,  ij\neq 11,44)$ $\triangle_{1,11}=\triangle_{1,44}=15.0g$, $\triangle _{2,11}=\triangle _{2,44}=15.2g$.} \end{figure}

To confirm the validity of all our above arguments, we numerically simulate the dynamics governed by the derived effective model in Eq. (12), and compare it to the dynamics governed by the full Hamiltonian
\begin{eqnarray}
H_f &=& \sum_{j,k=1}^N\{\omega_{f}|f\rangle _{jk}\langle f|+\omega_{e}|e\rangle _{jk}\langle e|+\sum[\omega_{c,jk}a_{jk}a_{jk} \nonumber\\
&&\ +g_{jk}a_{jk}|e\rangle _{jk}\langle g|+\Omega _{jk}|e\rangle _{jk}\langle f| e^{-i\omega_{l,jk}t}\nonumber\\
&&+\ va_{jk}a_{j+1k}^{+}+va_{jk}a_{jk+1}^{+}+h.c.]\},
\end{eqnarray}
where $\omega_{f}$ and $\omega_{e}$ are frequencies for the state $|f\rangle _{jk}$ and
$|e\rangle _{jk}$ ($|g\rangle _{jk}$ is assumed to be null energy level), and
$\omega_{c,jk}$ and $\omega_{l,jk}$ are frequencies for the cavity and
classical fields. We consider the case with $N=4$, and set the parameters in the following way : $v=1.5g$, $\Omega _{11}=\Omega _{44}=1.0g$, $\Omega _{ij}=0 (i,j=1,2,3,4,  ij\neq 11,44)$ $\triangle_{1,11}=\triangle_{1,44}=15.0g$, $\triangle _{2,11}=\triangle _{2,44}=15.2g$.
The validity of the effective model is numerically simulated by taking the evolution of the occupation probability $P=\left| \left\langle \psi |\psi(t) \right\rangle \right| ^2$ of the state $\left| f_{_{11}}g_{_{12}}\cdots g_{jk}\cdots g_{44}\right\rangle $ as an example, while assuming initially the atoms are in state $|\psi\rangle$ and all the cavity modes are in vacuum state.
FIG. 3 illustrates the numerical results obtained
from both the effective (red-dashed line) and full ( blue-solid line) Hamiltonians. Discrepancies between the two curves are due to higher terms for the parameters $\varsigma_1$ and $\chi_1$. Even now, it is obvious that the effective model is valid, its deviation can be made to be smaller as soon as the relative parameters are appropriately fixed.
 In Fig.4, the optimal fidelity of obtained state $\left| \Psi \right\rangle =\frac 1{\sqrt{2}}(\left| f_{_{11}}g_{_{12}}\cdots g_{jk}\cdots g_{44}\right\rangle +\left| g_{_{11}}\cdots g_{jk}\cdots g_{34}f_{44}\right\rangle )$
  is plotted with its initial state $\left| f_{_{11}}g_{_{12}}\cdots g_{jk}\cdots g_{44}\right\rangle$ versus time under the condition $v=1.5g$, $\Omega _{11}=\Omega _{44}=1.0g$, $\Omega _{ij}=0$ $(i,j=1,2,3,4,  ij\neq 11,44)$ $\triangle_{1,11}=\triangle_{1,44}=15.0g$, $\triangle _{2,11}=\triangle _{2,44}=15.2g$.  Fig.4
 shows that the maximally entangled state between the two atom with a fidelity higher than 0.99 can be obtained at time $t\simeq 100.0/g $s.

\begin{figure}
\centering
\includegraphics[width=3.6in]{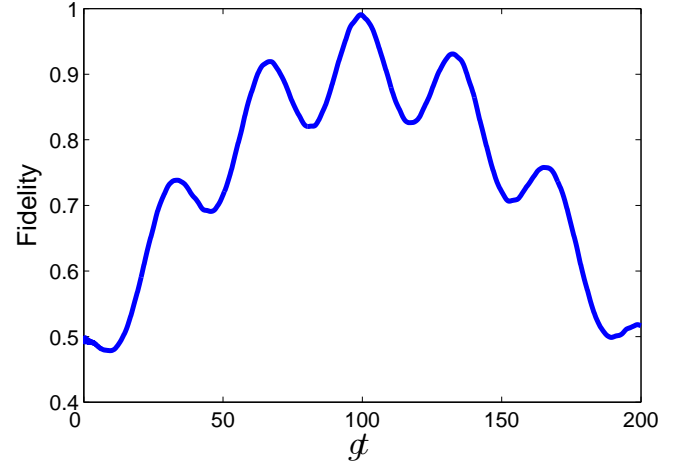}
 \caption{(Color online) The Fidelity of the state $\left| \Psi \right\rangle =\frac 1{\sqrt{2}}(\left| f_{_{11}}g_{_{12}}\cdots g_{jk}\cdots g_{44}\right\rangle +\left| g_{_{11}}\cdots g_{jk}\cdots g_{34}f_{44}\right\rangle )$ with its
 initial state $\left| f_{_{11}}g_{_{12}}\cdots g_{jk}\cdots g_{44}\right\rangle$ versus time. The maximally
 entangled state between the two atom state with a fidelity
 higher than 0.99 can be obtained at time $t\simeq 100.0/g $ s. The relative
 parameters are set the same as those in FIG3.}
 \end{figure}

It is necessary to give a brief discussion of the experimental feasibility of the proposed scheme. The proposed scheme can be realized in solid-state qubit trapped in a 2D array of  superconducting cavity system. In this system, the  superconducting cavity can be strongly coupled to the solid-state qubits such as Cooper pair boxes (CPB) and quantum dots (QD), and the microwave photons have small loss rates.  For $N=4$ and for the parameters introduced in FIG. (3), we get
$\chi_{jkpq}=\sum_{m,n}\frac 12\lambda _{11}\lambda _{22}(
\frac 1{\triangle _{1,11}-\omega _{mn}-\triangle _{2,11}}+\frac
1{\triangle
_{1,22}-\omega _{mn}-\triangle _{2,22}})e^{-i[2\frac{m\pi }N+2\frac{n\pi }N%
]}=7.63\times10^{-4}g$, and the time needed to complete the
entangling operation between qubit 11 and qubit 44 is $t=1.03\times10^{3}/g.$ The
probability that the atoms undergo a transition to the excited state
due to the off-resonant interaction with the classical fields is
$p_1=\frac{2\Omega ^2}{(\Delta _{2,11} )^2}=3.2\times10^{-3}$. Meanwhile,
the probability that the field modes are
excited due to off-resonant Raman couplings is $p_2=\sum_{jkmn}\frac{(\lambda _{jk})^2}{(\triangle _{1jk}+\omega _{mn}-\triangle _{2jk})^2}=6.98\times 10^{-3}$. The effective decoherence rates due to the
atomic spontaneous emission and the field decay are $\gamma
_e=p_1\gamma $ and $\kappa _e=p_2\kappa ,$ where $\gamma $ and
$\kappa $ are the decay rates for the atomic excited state and the
field modes, respectively. The parameter in a strongly coupled single quantum
dot-cavity system reported in Ref.\cite{Hennessy} is $\kappa \sim 5\times10^{-3}g($$\kappa =\frac g{1800})$, and
$\gamma  \sim 3\times10^{-3}g ($$\gamma =\frac g{300}).$  This leads to an entanglement fidelity with $F\simeq
1-(\gamma _e+\kappa _e)t\simeq 95\%.$

In conclusion, we propose a theoretical scheme to realize the
coherent coupling between two arbitrary qubits in a 2D quantum network.
We consider the scheme in a 2D coupled cavity system. In such a case, the
pairing off-resonant Raman transitions of distant atoms, induced
by the cavity modes and external fields, can lead to selective
coupling between two arbitrary atoms trapped in separated cavities;
quantum gates between any pair of qubits and parallel two-qubit operations
in the network can be performed. Moreover, the scheme can also be applied to N-V ensemble-based quantum network \cite{Benjiamin,Yangwangli}.

This work is supported from the Natural Science
Foundation of Fujian Province under Grant No. 2012J01010,
the Doctoral Foundation of the Ministry of Education of China
under Grant No.20093514110009.

\end{document}